\documentclass[%
 reprint, 
 amsmath,amssymb,
 aps,
]{revtex4-2}

\usepackage{graphicx}
\usepackage{dcolumn}
\usepackage{bm}
\usepackage{float}%
\usepackage{here}%
\usepackage{xspace}
\usepackage{mathtools}
\usepackage{tabularx}

\usepackage[dvipsnames]{xcolor}
\usepackage{stackengine}

\usepackage{bbm}

\newcommand{\add}[1]{{\color{black}#1}}
\newcommand{\addr}[1]{{\color{black}#1}}

\newcommand{\ER}{Erd\H{o}s-R\'{e}nyi }

\newcommand{\mo}{{\widetilde{m}}}

\newcommand{\qtrue}{{\bar{q}}}
\newcommand{\fa}{{\Theta}}
\newcommand{\namelambda}{accommodation threshold\xspace}
\newcommand{\thetamax}{\overline{\theta}}

\newcommand{\papertitle}{Social contagion induced by uncertain information}

\begin{document}

\title{\papertitle}%

\author{Teruyoshi Kobayashi}%
 \email{kobayashi@econ.kobe-u.ac.jp}
\affiliation{%
  Department of Economics, \\ Center for Computational Social Science, Kobe University, Kobe, Japan
}%

\date{\today}

\begin{abstract}
Information and individual activities often spread globally through the network of social ties. 
While social contagion phenomena have been extensively studied within the framework of threshold models, it is common to make an assumption that may be violated in reality: each individual can observe the neighbors' states without error.
Here, we analyze the dynamics of global cascades under uncertainty in an otherwise standard threshold model.
Each individual uses statistical inference to estimate the probability distribution of the number of active neighbors when deciding whether to be active, which gives a probabilistic threshold rule.
Unlike the deterministic threshold model, the spreading process is generally non-monotonic, as the inferred distribution of neighbors' states may be updated as a new signal arrives.
We find that social contagion may occur as a self-fulfilling event in that misperception may trigger a cascade in regions where cascades would never occur under certainty.
\end{abstract}

\maketitle

\section{Introduction}

 The spreading of information, opinions, and other social activities through social ties is collectively called social contagion~\cite{burt1987social,ugander2012structural,christakis2013social,ruan2015kinetics}.   
 The dynamical processes of social contagion on networks have been extensively studied over the past decades, and a wide variety of cascade models have been developed to describe various cascading behaviors in different social/economic contexts, from cultural fads to financial crises~\cite{morris2000contagion,Watts2002,Watts2007,GaiKapadia2010,Centola2010Science,Brummitt2012_PNAS,Cont2013,Brummitt2015PRE,kobayashi2015trend,Nematzadeh2014,burkholz2016damage,burkholz2018explicit,
 Guilbeault2018,caccioli2018review,unicomb2018threshold,unicomb2019reentrant,unicomb2021dynamics}.
 
 A workhorse framework is the class of threshold models in which individual behavior is influenced by local neighbors if the fraction of ``active'' neighbors exceeds a certain threshold~\cite{Granovetter1978,Watts2002}. 
 In the threshold models, it is presumed that individuals can observe the neighbors' states and decide whether to be active without error according to a deterministic threshold rule.
 However, the observability of the states of others is not necessarily perfect in many real-world contexts~\cite{Guilbeault2018}. 
 For example, it would be difficult to accurately observe the number of friends who believe a certain rumor.
 One might like to subscribe to a YouTube channel if most friends are already subscribed, but the status of subscription may be unavailable. 
 
 In environments where the neighbors' states are uncertain, individuals need to infer the number of active neighbors before making decisions.
 This could change the fundamental mechanics of social contagion since individual decisions would be based on statistical inference rather than direct observation. 
Here, we develop a threshold model of cascades in the presence of observation uncertainty.
Instead of assuming that the true number of active neighbors is known, each individual is assumed to observe a noisy signal from which the probability distribution of the neighbors' states is inferred.
Then, each individual decides whether to be active or not based on a probabilistic threshold rule.
In this uncertain environment, an individual's misperception could trigger a global cascade, even in regions where cascades would never occur without uncertainty.

\section{Noisy information and statistical inference}

\add{
\subsection{Deterministic threshold rule}

Before we get into the model with uncertain information, let us briefly describe the standard binary-state model of social contagion, where inactive individuals decide whether to be active in a deterministic manner.
Our model follows from the Watts model of global cascades~\cite{Watts2002}, in which individuals having $k$ neighbors become active if
\begin{align}
    \frac{m}{k}>\theta, 
    \label{eq:watts_threshold_rule}
\end{align}
where $m$ denotes the number of active neighbors, and $\theta\in(0,1)$ is an activation threshold.
The individuals remain inactive if $m/k\leq \theta$.

The condition~\eqref{eq:watts_threshold_rule} represents a \emph{deterministic} threshold rule as it is implicitly assumed that the true number of active neighbors $m$ can be observed without error. 
\emph{All} the individuals that satisfy the deterministic threshold condition surely become active.
If the threshold $\theta$ is smaller than a certain critical value, a small number of ``seed nodes'' may trigger a global contagion in which a large fraction of nodes in the network become active.
It is known that such a binary-state contagion process is monotonic in that active nodes cannot revert to the inactive state~\cite{Gleeson2007,kobayashi2022EconTheo}. 
The steady state and dynamical path of the total fraction of active nodes are often calculated using a message-passing method, which is analytically tractable and highly accurate for the class of deterministic threshold models~\cite{Gleeson2007,Brummitt2015PRE,gleeson2018message,kobayashi2022EconTheo}.
}

\subsection{Probabilistic threshold rule}
\begin{figure}[tb]
    \centering
    \includegraphics[width=6cm]{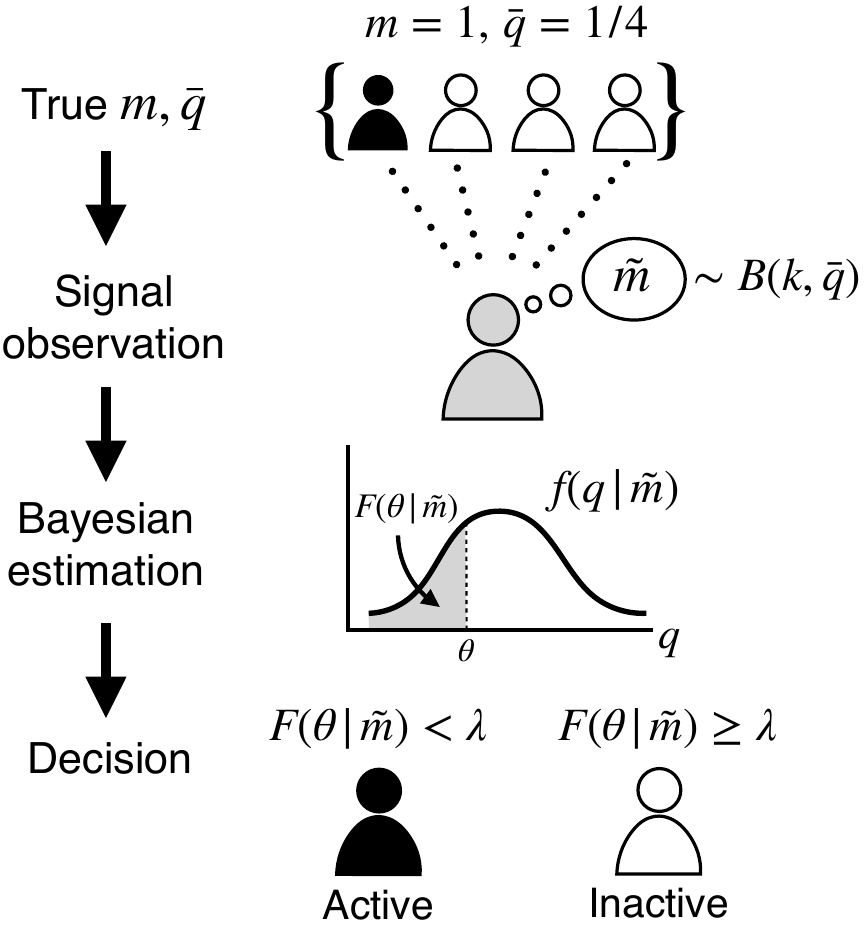}
    \caption{
    \add{Schematic of Bayesian inference and node activation.
    An individual does not have complete information about the true number of active neighbors $m$ and the fraction of active neighbors $\qtrue$.
    The individual receives a signal $\mo$ that follows a binomial distribution $B(k,\qtrue)$.
   Conditional on the realized signal, the distribution of the true fraction of active neighbors (denoted by $f(q|\mo)$) is inferred.
    The individual is activated if $F(\theta|\mo)<\lambda$, where $F(\theta|\mo)=\int_0^\theta f(q|\mo)dq$.
    }
    }
    \label{fig:schematic_estimation}
\end{figure}

\add{
Now consider a situation in which the true number of active neighbors $m$ is unobservable.
The state of each node is still either active or inactive, but the node states are unknown to their neighbors, while each node knows the number of neighbors (i.e., degree) $k$.
Each node observes a noisy signal of $m$, denoted by $\mo$.
$\mo$ is distributed in $[0,k]$, following a binomial distribution with mean $m$: $\mo\sim B(k,\qtrue)$, where the unobservable probability $\qtrue\coloneqq  m/k$ is the true fraction of active neighbors.
That is, a signal $\mo$ is interpreted as the number of ``successes'' in $k$ trials with a success probability of $\qtrue$.
For a given true fraction $\qtrue$, a signal $\mo$ is drawn independently of the history of node states, while $\qtrue$ generally changes over time once a cascade occurs.
Thus, the distribution of $\mo$ shifts as $\qtrue$ changes. 

When deciding whether to be active, each node infers the distribution of the true fraction $\qtrue$ based on the observed signal $\mo$.
The posterior distribution for $\qtrue$, denoted by $f(q|\mo)$, is obtained from the following Bayes' rule:
\begin{align}
 f(q|\mo) = \frac{f(\mo|q)f(q)}{\int_{0}^1 f(\mo|q)f(q)dq},
\label{eq:bayes_rule}
\end{align}
where $f(\mo|q) = \binom{k}{\mo}q^\mo (1-q)^{k-\mo}$, and $f(q)$ is a prior distribution.

We specify that a node is activated if and only if the following probabilistic threshold condition is satisfied: 
\begin{align}
    {\rm Prob}\left(q >\theta\right|\mo) > 1-\lambda,
    \label{eq:prob_threshold}
\end{align}
where $\lambda\in(0,1)$ represents an \emph{\namelambda}. A node is activated if the probability of the fraction of active neighbors exceeding the threshold $\theta$ is larger than $1-\lambda$.   
The higher the \namelambda $\lambda$, the more likely the node will be activated.
}
Using the posterior distribution $f(q|\mo)$, the probabilistic threshold condition \eqref{eq:prob_threshold} is rewritten as
\begin{align}
    \int_0^\theta f(q|\mo)dq\coloneqq  {F}(\theta| \mo ) < \lambda,
\label{eq:threshold_F}
\end{align}
where $F$ is the cumulative density of the posterior distribution.
A schematic of an individual's decision-making based on statistical inference is presented in Fig.~\ref{fig:schematic_estimation}.

To facilitate the analysis, we assume that the prior $f(q)$ is given by a beta distribution, ${\rm Beta}(\alpha,\beta)$.
Because the beta distribution is a conjugate for the binomial distribution, the posterior distribution $f(q|\mo)$ in Eq.~\eqref{eq:bayes_rule} leads to ${\rm Beta}(\alpha+\mo,\beta+k-\mo)$.


\subsection{Degree of uncertainty} \label{sec:degree_uncertainty}
 
 \add{
 We specify the prior distribution $f(q)$ in such a way that the degree of uncertainty in the number of active neighbors is properly measured by the variance of $q$.
 For this purpose, $f(q)$ must be an ``unbiased'' distribution with the mean equal to the true value $\qtrue$ because otherwise, node states would be affected by biased estimation as well as the effect of uncertainty.
 If we use a ``biased'' prior such that $\mathbb{E}(q)\ne \qtrue$, then a global cascade could occur due to the biased belief even if no node in the network is actually active (e.g., if we use the uniform prior ${\rm Beta}(1,1)$, we would have $\mathbb{E}(q)=\int_0^1qdq= 1/2$ and $\mathbb{E}(q|\mo=0)=1/(2+k)$).
 To avoid this, we will make sure that $\mathbb{E}(q)=\qtrue$ and $\mathbb{E}(q|\mo=0)=0$.
 
Since an unbiased beta prior must have mean $\alpha/(\alpha+\beta)=\qtrue$, $\alpha$ is set at $\alpha=\frac{\qtrue}{1-\qtrue}\beta$ for $\qtrue\in(0,1)$ and $\beta>0$.
Then, the mean and variance of the prior distribution lead to 
\begin{align}
\mathbb{E}(q) &= \qtrue,\\
{\rm Var}(q)&= \frac{\qtrue(1-\qtrue)^2}{1-\qtrue+\beta},
\end{align}
which implies that the degree of uncertainty decreases monotonically with $\beta$, suggesting that $\beta$ can be used as a proxy for the observability of $\qtrue$ (Fig.~2a).
Note that as $\qtrue\to 0$ or $1$, ${\rm var}(q)$ will vanish, and individuals can observe $\qtrue$ accurately.
Therefore, regardless of $\beta$, uncertainty will disappear in the special cases where no neighbor or all neighbors are active.
For this reason, nodes having no active (no inactive) neighbors always become inactive (active). 
This allows us to eliminate the possibility of a cascade occurring without active nodes.
}

\begin{figure}[tb]  
    \centering
    \includegraphics[width=8.6cm]{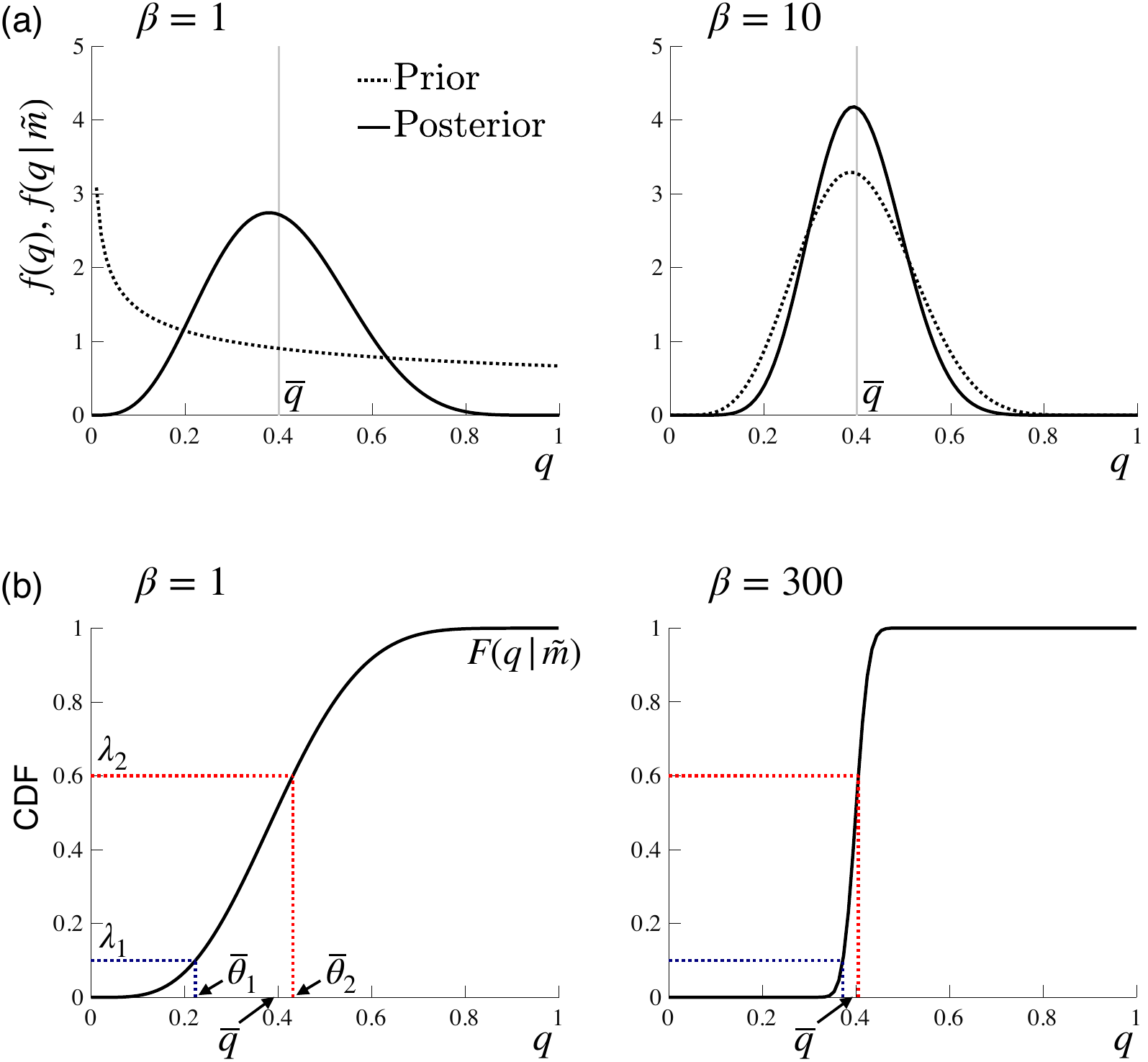}
    \add{
    \caption{Prior and posterior distributions.
    (a) Prior (dotted) and posterior (black solid) distributions are more concentrated around the true fraction of active nodes $\qtrue$ (vertical dotted line) when $\beta$ is larger.
    (b) CDF of  the posterior distribution, $F(q|\mo)$.
    $\thetamax_1$ ($\thetamax_2$) denotes the threshold of $\theta$ below which the node is activated for a given accommodation level $\lambda_1$ ($\lambda_2$).  
    We have $\theta_1\approx \theta_2$ when $\beta$ is sufficiently large.
    We set $k=10$, $\qtrue=0.4$, and $\mo=4$.}
    }
\label{fig:f_dist}
\end{figure}

\add{
Because the prior mean is unbiased, the model is equivalent to the classic threshold model of Watts~\cite{Watts2002} in the limit of large $\beta$, in which $m/k$ is revealed without error.
This results in equivalence between the prior and posterior means and variances.:
}
\begin{align}
    \lim_{\beta\to\infty}{\mathbb E}(q|\mo) &= \lim_{\beta\to\infty}\frac{\mo+\frac{\qtrue\beta}{1-\qtrue}}{\frac{\qtrue\beta}{1-\qtrue} + \beta+ k}  =\qtrue,\\
    \lim_{\beta\to\infty}{\rm Var}(q|\mo) &= \lim_{\beta\to\infty}\frac{\left(\frac{\qtrue\beta}{1-\qtrue}+\mo\right)(\beta+k-\mo)}{\left(\frac{\beta}{1-\qtrue}+k\right)^2\left(\frac{\beta}{1-\qtrue}+k+1\right)} =0.
\end{align}
When there is no uncertainty (i.e., ${\rm Var}(q|\mo)=0$), we have $F(q|\mo)=0$ for $q<\qtrue$, and $F(q|\mo)=1$ for $q\geq\qtrue$.
Then, the threshold condition~\eqref{eq:threshold_F} is satisfied if and only if $\theta < \qtrue=m/k$ for any $\lambda\in(0,1)$, which recovers the threshold rule \`{a} la Watts~\cite{Watts2002}.
\add{
Fig.~2b illustrates the correspondence between $\lambda$ and the range of $\theta$ within which the node is activated.
For a given accommodation level $\lambda$, there is a unique value $\thetamax$ such that $\lambda=F(\thetamax|\mo)$, or $\thetamax = F^{-1}(\lambda|\mo)$.
That is, $\thetamax$ is the threshold of $\theta$ below which the node is activated.
While $\thetamax$ is generally positively correlated with $\lambda$, the correlation weakens as $\beta$ increases (i.e., as uncertainty decreases). 
$\thetamax$ is virtually independent of $\lambda$ for a sufficiently large $\beta$ (Fig.~2b, \emph{right}).

In the following analysis, we consider a situation in which each node uses an unbiased prior ${\rm Beta}(\alpha(\qtrue),\beta)$ when updating its state, where $\alpha(\qtrue)=\qtrue\beta/(1-\qtrue)$.  
The prior parameter $\alpha$ thus evolves with $\qtrue$, while $\beta$ is constant and treated as a tuning parameter that controls the degree of uncertainty.
In Sec.~\ref{sec:sequential_update}, we also consider an alternative Bayesian updating scheme in which prior distribution is updated sequentially based on past observations.
}  
 
\section{Analysis}

\subsection{Spreading dynamics}

We describe the spreading process using the approximate master equation (AME) method~\cite{gleeson2011high,gleeson2013binary,fennell2019multistate}. 
Although the message-passing method has been extensively used in analyzing threshold models with certainty, it would not be suitable here as the spreading process is not necessarily monotonic. 
Throughout the paper, we consider networks that are locally tree-like and have only a negligible number of local cycles.
In the AME approximation, it is assumed that the state transition rate of a neighbor of a node is independent of the states of the other neighbors of the node~\cite{fennell2019multistate}. 

Let $s\in \{0,1\}$ ($=\{{\rm inactive},{\rm active}\})$ be the state of a node, and
$\rho^s_{k,m}$ be the fraction of $k$-degree nodes in state $s$ having $m$ active neighbors (or equivalently, having $k-m$ inactive neighbors), where $\sum_{m=0}^k \rho_{k,m}^s=\rho_k^s$ is the fraction of $k$-degree nodes in state $s$, and the expected total fraction of nodes in state $s$ is given by $\rho^s = \sum_k p_k\sum_{m=0}^k \rho_{k,m}^s$.

Using the AME formalism~\cite{gleeson2011high,gleeson2013binary,fennell2019multistate}, we express the evolution of $\rho^{s}_{k,m}$ as follows (the notations are summarized in Table~\ref{tab:AME_notations}):
\addr{
\begin{align}
    \dot{\rho^{1}}_{k,m} =\; & -\overbrace{R_{k,m}(0)\rho_{k,m}^{1}}^{\substack{\text{Active $(k,m)$ nodes}\\{\text{ become inactive}}}}
    -\overbrace{m\phi(1\to 0)\rho_{k,m}^{1}}^{\substack{\text{Active neighbors of an active}\\ \text{$(k,m)$ node become inactive}}} \notag \\
     & + \underbrace{R_{k,m}(1)\rho_{k,m}^{0}}_{\substack{\text{Inactive $(k,m)$ nodes}\\ \text{become active}}} 
     - \underbrace{(k-m)\phi(0\to 1)\rho_{k,m}^1}_{\substack{\text{Inactive neighbors of an active} \\ \text{$(k,m)$ node become active}}} \notag \\ 
     &+  \underbrace{(k-m+1)\phi(0\to 1)\rho_{k,m-1}^{1}}_{\substack{\text{Neighbors of an active $(k,m\!-\!1)$}\\ \text{node become active}}} \notag\\
     &+  \underbrace{(m+1)\phi(1\to 0)\rho_{k,m+1}^{1}}_{\substack{\text{Neighbors of an active $(k,m\!+\!1)$}\\ \text{node become inactive}}}, \label{eq:AME_eqs}\\
    \dot{\rho^{0}}_{k,m} =\; &-\dot{\rho^{1}}_{k,m}, 
\label{eq:AME_eqs_s0}
\end{align}
}
\begin{table}[tb]
 
    \centering
        \caption{Notations in the approximate master equation \eqref{eq:AME_eqs}.}
    \begin{tabularx}{.48\textwidth}{lX}
    \hline
     Notation$\;\;$ & Description \\
     \hline
     $k$ & Degree of a node \\
     $m$ & Number of active neighbors of a node \\
     $\mo$ & Signal of $m$ \\
     ${\rho}_{k,m}^s$    & Fraction of nodes in state~$s$ belonging to the $(k,m)$ class \\
     $R_{k,m}(s^\prime)$     & The rate at which nodes in the $(k,m)$ class change their states to $s^\prime$  \\
     $\phi(s\to s^\prime)$ & The rate at which a neighbor of an active node changes its state from $s$ to $s^\prime$ \\
     $p_k$  &  Probability of a node having degree $k$\\
     \hline
    \end{tabularx}
    \label{tab:AME_notations}
    
\end{table}
where $\phi(s\to s^\prime)$ denotes the probability that a neighbor of an active node changes its state from $s$ to $s^\prime$: 
\begin{align}
    \phi(s\to s^\prime) =& \frac{\overbrace{\sum_k p_k\sum_{m=1}^k  m{\rho}^{s}_{k,m}R_{k,m}(s^\prime)}^{\substack{\text{Expected no. of edges changing}\\ \text{from $(1)$-$(s)$ to $(1)$-$(s^\prime)$}}}}{\underbrace{\sum_k p_k\sum_{m=1}^k  m \rho_{k,m}^{s}}_{\text{Expected no. of $(1)$-$(s)$ edges}}}.
    \label{eq:phi}
\end{align}
$p_k$ denotes the degree distribution, and the stochastic response function $R_{k,m}(s^\prime)$ describes the rate at which individuals in the $(k,m)$ class change their state to $s^\prime$:
\begin{align}
    & R_{k,m}(s^\prime) &\notag \\
    &= \begin{cases}
     {\rm Prob}\left[F(\theta|\mo) < \lambda\, \big|\, (k,m)\right] & {\text{ if }}\; s' = 1  \\ 
    1-{\rm Prob}\left[F(\theta|\mo) < \lambda\, \big|\, (k,m)\right] & {\text{ if }}\; s' = 0,
    \end{cases}
    \label{eq:response_func}
\end{align}
where
\begin{align}
&  {\rm Prob}\left[F(\theta|\mo) < \lambda\, \big|\, (k,m)\right]  \notag \\
&= \sum_{\mo=0}^k\binom{k}{\mo}\qtrue^\mo(1-\qtrue)^{k-\mo}\,\mathbbm{1}_{F(\theta|\mo) < \lambda}.
\label{eq:F_in_R}
\end{align}
$\mathbbm{1}_A$ is the indicator function that takes $1$ if condition $A$ is satisfied and $0$ otherwise.
It should be noted that while the response of each individual is deterministic for a given $\mo$ (Eq.~\ref{eq:threshold_F}), it is stochastic ex ante since $\mo$ is a random variable following a binomial distribution $B(k,\bar{q})$.
Eq.~\eqref{eq:response_func} also suggests that the probability that a node's state becomes $s^\prime$ is independent of the current state of the node.
Therefore, the next state of a node is determined only by the signal and threshold values, so the state of a node can change from active to inactive.
This possibility of state reversal is absent in the standard threshold model without uncertainty.  
In Sec.~\ref{sec:state_reversal}, we will examine the extent to which nodes revert their states during a cascade process.

 In Eq.~\eqref{eq:AME_eqs}, there are four factors that will change $\rho_{k,m}^1$. Active nodes with degree~$k$ will \emph{leave} the $(k,m)$ class if \emph{i}) their state changes from $1$ to $0$ (the first term) or \emph{ii}) the number of active neighbors changes from $m$ to $m^\prime (\neq m)$ \addr{(the second and fourth terms)}. 
On the other hand, nodes with degree~$k$ will \emph{enter} the $(k,m)$ class if \emph{iii}) inactive nodes having $m$ active neighbors newly become active (the third term) or \emph{iv}) the number of active neighbors of active nodes shifts from $m^\prime(\neq m)$ to $m$ \addr{(the fifth and sixth terms)}.

The denominator of Eq.~\eqref{eq:phi} represents the expected number of edges between active nodes (i.e., nodes in state~$1$) and nodes in state~$s\in\{0,1\}$.
We call such edges $(1)$--$(s)$ edges.
Since the expected number of $(1)$--$(s)$ edges that change to $(1)$--$(s^\prime)$ in an infinitesimal interval $dt$ is given as $\sum_k p_k \sum_{m=1}^k m\rho_{k,m}^s R_{k,m}(s^\prime)dt$, the probability of a $(1)$--$(s)$ edge shifting to a $(1)$--$(s^\prime)$ edge, denoted by $\phi(s\to s^\prime)dt$, is obtained as the ratio of the two, leading to Eq.~\eqref{eq:phi}.
The AME solution is calculated using the Matlab codes provided in~\cite{FennelCode}.

  \subsection{Cascade conditions}
 \add{
  The AME method accurately predicts the cascade size $\rho^1$, but the calculation can be computationally expensive when there is a large number of differential equations. Note that in the AMEs, we have equations for $\rho_{k,m}^0$ and $\rho_{k,m}^1$ for each of the combinations of $\{(k,m)\}$. 
  An alternative, though less accurate, method is the mean-field (MF) approximation, for which it is assumed that the state, as opposed to the transition rate, of a neighbor is independent of the states of the other neighbors.
  }
  The MF equation is given by
  \begin{align}
    \dot{\rho^1} = 
    &-\rho^1\sum_k p_k\sum_{m=0}^k\binom{k}{m}\left(\rho^1\right)^m\left(1-\rho^1\right)^{k-m} \notag\\
     &\qquad\times {\rm Prob}\left[F(\theta|\mo) \geq \lambda\, \big|\, (k,m)\right] \notag\\
     &+ (1-\rho^1)\sum_k p_k\sum_{m=0}^k\binom{k}{m}\left(\rho^1\right)^m\left(1-\rho^1\right)^{k-m} \notag \\
    &\qquad\times {\rm Prob}\left[F(\theta|\mo) < \lambda\, \big|\, (k,m)\right] \label{eq:MF_eq_pre}\\
     =& -\rho^1 + \sum_k p_k\sum_{m=0}^k\binom{k}{m}\left(\rho^1\right)^m\left(1-\rho^1\right)^{k-m}\notag\\
     &\qquad\times {\rm Prob}\left[F(\theta|\mo) < \lambda\, \big|\, (k,m)\right],
 \label{eq:MF_eq}
\end{align}
where the first (second) term in Eq.~\eqref{eq:MF_eq_pre} represents the fraction of nodes that change the state from $s=1$ to $0$ (from $s=0$ to $1$).

\begin{figure}[tb]
    \centering
    \includegraphics[width=8.6cm]{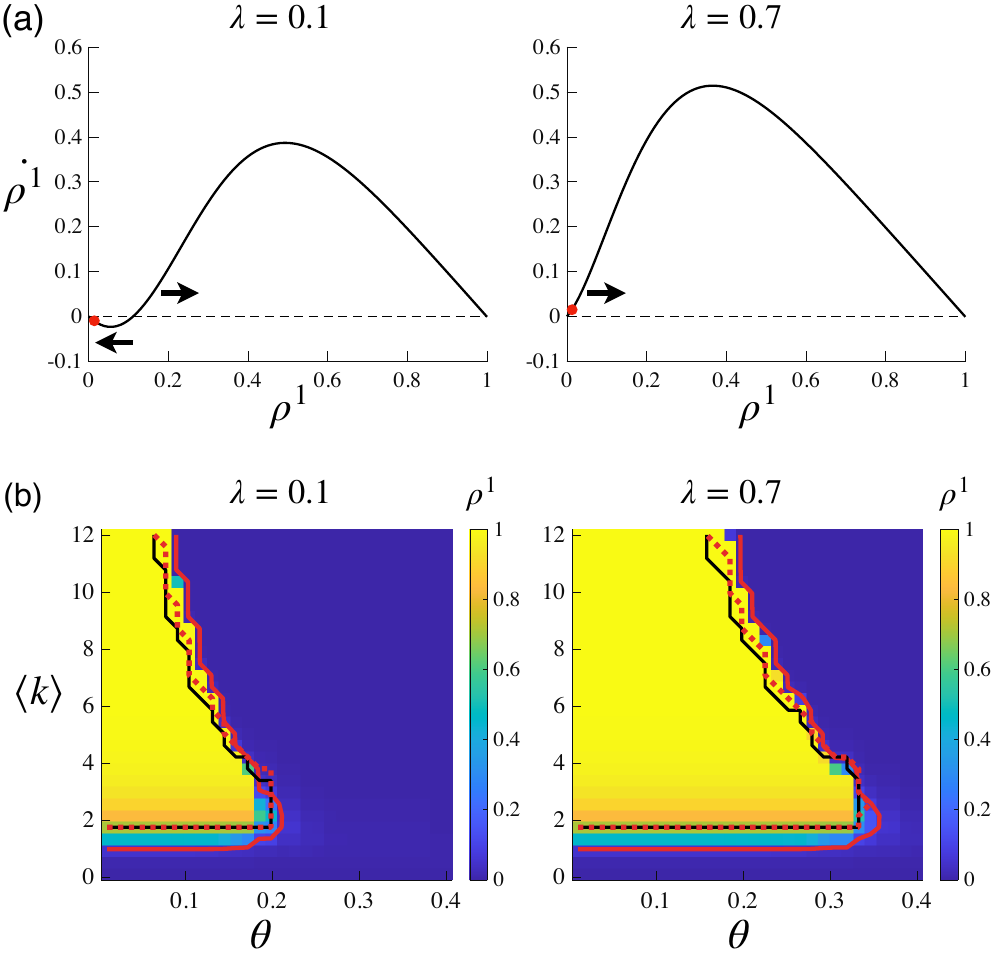}
    \caption{Phase diagram and cascade region. 
    (a) Phase diagram based on the MF equation~\eqref{eq:MF_eq} with $\lambda=0.1$ and $\lambda=0.7$. 
    Red circle indicates $\rho^1=\rho_0=0.01$. $\theta=0.18$, $\beta=20$, and $\langle k\rangle=8$.
    (b) Cascade region obtained by simulation (color) and the cascade conditions (lines). 
    The areas surrounded by red solid, red dashed and black solid lines denote the cascade regions indicated by the AMEs (i.e., $\rho_1(T)>0.1$), the sign condition \eqref{eq:cascade_condition_full}, and the first-order cascade condition \eqref{eq:cascade_condition_foc}, respectively.
    We set $N=10^5$ and $\beta=20$.
    }
    \label{fig:phase_diagram}
\end{figure}

Phase diagrams for the case of \ER networks are drawn in Fig.~\ref{fig:phase_diagram}a using the MF equation~\eqref{eq:MF_eq}.
 The \add{\namelambda}  $\lambda$ considerably affects the shape of $\dot{\rho^1}$ and the possibility of a global cascade.
For a small $\rho^1$, $\dot{\rho^1}$ is likely to take a negative value when $\lambda$ is small, in which $\rho^1$ converges to $0$.
In contrast, $\dot{\rho^1}$ is likely to take a positive value when $\lambda$ is large, in which $\rho^1$ diverges from $0$, indicating the onset of a global cascade.
Therefore, a sign condition for a global cascade to occur at $\rho^1=\rho_0\coloneqq \rho^1|_{t=0}>0$ is given by $\dot{\rho^1}|_{\rho^1=\rho_0}>0$, or
\begin{align}
   &\sum_k p_k\sum_{m=0}^k\binom{k}{m}\left(\rho_0\right)^m\left(1-\rho_0\right)^{k-m} \notag\\
    &\qquad\times {\rm Prob}\left[F(\theta|\mo) < \lambda \,\big|\, (k,m)\right]   > \rho_0.
\label{eq:cascade_condition_full}
\end{align}
In the limit of $\rho^1\to 0$, we have $\dot{\rho^1}\to 0$ from Eq.~\eqref{eq:MF_eq}, in which case the sign condition \eqref{eq:cascade_condition_full} would not be useful.
Instead, we could use another cascade condition, called the \emph{first-order cascade condition}: $\partial \dot{\rho^1}/\partial \rho^1|_{\rho^1=0}>0$, or
\begin{align}
    \sum_k p_k k \sum_{\mo=0}^k\binom{k}{\mo}\left(\frac{1}{k}\right)^\mo\left(1-\frac{1}{k}\right)^{k-\mo}\mathbbm{1}_{F(\theta|\mo)< \lambda}>1,
    \label{eq:cascade_condition_foc}
\end{align}
\add{
where the parameter $\alpha$ for the posterior distribution $F$ (i.e., beta distribution) is given by $\alpha =\beta/(k-1)$, since the terms associated with $m\ne 1$ are all zero. 
}
The first-order condition~\eqref{eq:cascade_condition_foc} can also be expressed as
\begin{align}
    \mathbb{E}_k\left[ k\cdot {\rm Prob}[F(\theta|\mo) < \lambda \,\big| \, (k,1)]\, \right]>1.
    \label{eq:cascade_condition_expected}
\end{align}
The LHS of Eq.~\eqref{eq:cascade_condition_expected} represents a reproduction number: the expected number of nodes affected by an active neighbor. Thus, the condition indicates that global cascades can occur if, on average, more than one node will become active when there is initially only one active neighbor.
Eqs.~\eqref{eq:cascade_condition_foc} and \eqref{eq:cascade_condition_expected} can be viewed as the stochastic counterpart of the standard cascade condition under certainty~\cite{Watts2002,Gleeson2007,gleeson2018message,kobayashi2022EconTheo}.

A comparison between the simulated and analytical cascade regions is shown in Fig.~\ref{fig:phase_diagram}b.
The cascade region generally expands as the \namelambda  $\lambda$ increases, which is well captured by the analytical cascade conditions \eqref{eq:cascade_condition_full} and \eqref{eq:cascade_condition_foc} as well as the solution of the AMEs.

 \subsection{Role of misperception}
 
 Because the true number of active neighbors is generally unobservable, individuals' decisions may be misguided by noisy signals.
 The dynamical process of contagion would be purely deterministic if there were no observation errors (for a given network and a given set of seed nodes), but noisy signals may sometimes lead individuals to make ``wrong'' decisions; some nodes may become \emph{false active} if they overestimate $\qtrue$, or \emph{false inactive} if they underestimate $\qtrue$.
 The possibility of misperception at the individual level could affect the macroscopic outcome and result in a self-fulfilling social contagion.

\subsubsection{Uncertainty effect}

Let $\mathcal{C}(\rho_0)$ denote the cascade region that specifies the range of parameters within which the sign condition~\eqref{eq:cascade_condition_full} is satisfied for a given seed fraction $\rho_0>0$.
\begin{align}
    \mathcal{C}(\rho_0) =& \Bigl\{(\theta, \lambda,\langle k\rangle, \beta)\, \colon 
    \sum_k p_k\sum_{m=0}^k B_m^k(\rho_0) \notag\\
    &\qquad \times{\rm Prob}[F(\theta|\mo) < \lambda \,\big|\, (k,m)]   > \rho_0 \Bigr\},
\end{align}
where $B_m^k(\rho_0)=\binom{k}{m}\left(\rho_0\right)^m\left(1-\rho_0\right)^{k-m}$.
The cascade region under certainty, in which there are no false-active or false-inactive nodes, is given as a limiting case of $\beta\to\infty$.  
Let $\widetilde{\mathcal{C}}(\rho_0)\coloneqq\mathcal{C}(\rho_0)|_{\beta\to\infty}$ denote the cascade region under certainty.
If $\mathcal{C}\setminus\widetilde{\mathcal{C}}\neq \emptyset$, then there exists a nonempty cascade region in which global cascades are \emph{caused} by false-active nodes.
If $\widetilde{\mathcal{C}}\setminus\mathcal{C}\neq \emptyset$, by contrast, there exists a region in which global cascades are \emph{suppressed} due to the prevalence of false-inactive nodes.  

\begin{figure}[tb]
    \centering
    \includegraphics[width=8.6cm]{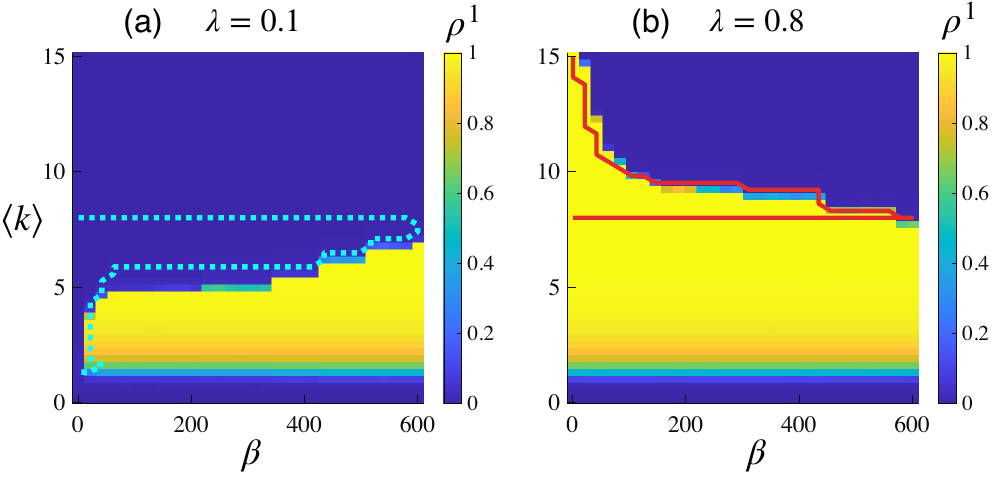}
    \caption{Shrinkage and expansion of cascade region induced by misperception. The heat maps represent the simulated fraction of active nodes for (a) $\lambda=0.1$ and (b) $\lambda=0.8$.  
    Area surrounded by blue-dotted (red solid) line represents the difference from the cascade region under certainty, $\tilde{\mathcal{C}}\setminus\mathcal{C}$ (${\mathcal{C}}\setminus\tilde{\mathcal{C}}$), indicated by the sign  condition~\eqref{eq:cascade_condition_full}.
    $N=10^5$, $\theta=0.18$, and $\rho_0=0.01$.
    Simulation results show the average over 100 runs.
    }
    \label{fig:misperception_heatmap}
\end{figure}

When $\lambda$ is small, the cascade region with uncertain information is smaller than that without uncertainty (Fig.~\ref{fig:misperception_heatmap}a). 
This suggests that false-inactive nodes (i.e., nodes that underestimate the fraction of active neighbors) contribute to inhibiting a cascade from spreading widely, compared to the standard model with complete information. 
In this regime, therefore, the cascade region expands as the degree of uncertainty decreases (i.e., as $\beta$ increases).
In contrast, when $\lambda$ is large, the cascade region becomes larger than that under certainty (Fig.~\ref{fig:misperception_heatmap}b).
This is a situation in which false-active nodes (i.e., nodes that overestimate the fraction of active neighbors) enhance  global cascades. 

\begin{figure}[tb]
    \centering
    \includegraphics[width=8.5cm]{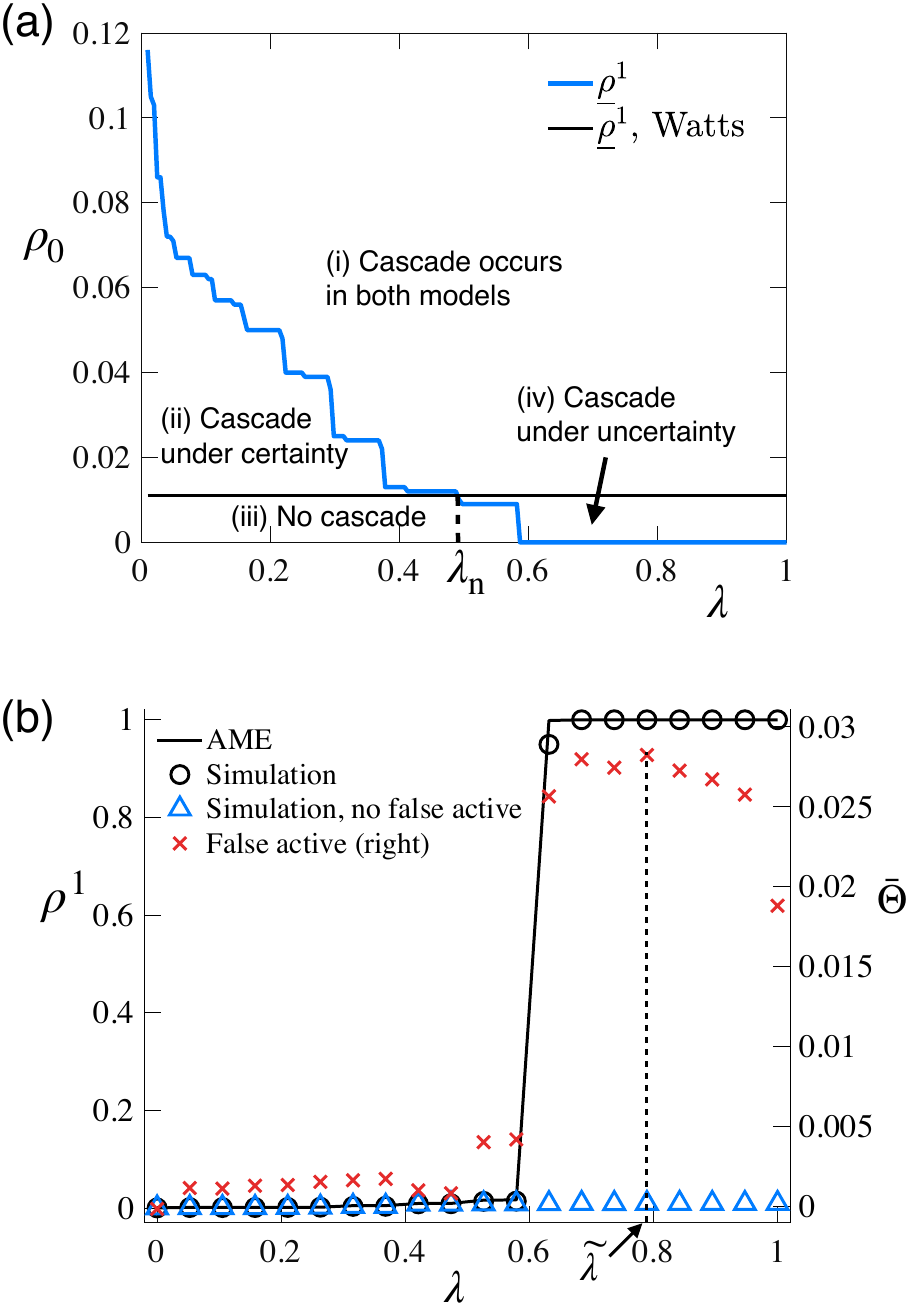}
    \caption{Effects of varying the \namelambda  $\lambda$. (a) The largest unstable solution of Eq.~\eqref{eq:MF_eq}
    (i.e., the threshold of $\rho_0$ above which a global cascade occurs). Blue and black lines respectively denote the values of $\underline{\rho}^1$ obtained by the models with and without uncertainty.
    \add{
    $\lambda_{\rm n}$ denotes the neutral level of accommodation at which the effect of uncertainty is offset.
    }
    (b) Theoretical (black line) and simulated (black circle) fraction of active nodes. 
    Red cross denotes the normalized cumulative fraction of false-active nodes, $\bar\Theta$ (right axis), which takes the maximum at $\widetilde{\lambda}$.
    Blue triangle shows the simulated fraction of active nodes that would be attained without false active nodes. 
    We set $N=10^5$, $\langle k\rangle =8$, $\theta=0.18$, $\beta=100$ and $\rho_0=0.01$.
    Simulation results show the average over 100 runs.
    }
    \label{fig:cumulative_false_active}
\end{figure}

\add{
Therefore, the effect of uncertainty varies depending on the individuals' accommodation level $\lambda$.
Fig.~\ref{fig:cumulative_false_active}a illustrates the minimum seed fraction, denoted by $\underline{\rho}^1$, needed to trigger a global cascade for a given $\lambda$.
Here, we obtain $\underline{\rho}^1$ as the largest unstable solution of Eq.~\eqref{eq:MF_eq}. 
As $\lambda$ increases, $\underline{\rho}^1$ decreases toward $0$ (blue line in Fig.~\ref{fig:cumulative_false_active}a) since more accommodative individuals can cause a cascade with a relatively smaller seed fraction.
 In the case of Fig.~\ref{fig:cumulative_false_active}a, we find that $\underline{\rho}^1$ is approximately $0.01$ in the model without uncertainty (i.e., black line), and there is a point at which the effect of uncertainty is offset (indicated by $\lambda_{\rm n}$). 
$\lambda_{\rm n}$ is regarded as the \emph{neutral level} of accommodation.

The $(\lambda,\rho_0)$ space in Fig.~\ref{fig:cumulative_false_active}a is divided into four regions (i)--(iv) according to the possibility of social contagion due to uncertainty. These are characterized as follows: (i) cascades occur regardless of uncertainty, (ii) cascades occur only if there is no uncertainty, (iii) cascades will never occur, and (iv) cascades cannot occur without uncertainty. 
The regions~(ii) and (iv) exhibit the two aforementioned contrastive uncertainty effects: in region~(ii), uncertainty inhibits cascades because individuals are conservative in the sense that $\lambda<\lambda_{\rm n}$.
In contrast, uncertainty enhances cascades in region~(iv) since individuals are accommodative enough to follow seemingly active neighbors.
}

\subsubsection{Self-fulfilling contagion}
\add{
In the absence of uncertainty, where there are no false active nodes other than seed nodes, nodes become active if and only if a sufficiently large fraction of their neighbors is truly active.
On the other hand, when the true number of active neighbors is unobservable, the decision to be active depends on the inferred distribution of the number of active neighbors, which raises the possibility that there may be many false active nodes. 
The presence of false active nodes will not be related to the likelihood of a global cascade as long as $\lambda$ is smaller than the neutral level.
However, when $\lambda$ is greater than the neutral level, where individuals are highly accommodative, false-active nodes can trigger a global cascade.
In the latter case, the cascade region will be larger than it would be without uncertainty. 
In the model with uncertainty, therefore, it is important to quantify to what extent the presence of uncertainty increases/decreases the likelihood of cascades.

In this section, we quantify the importance of false-active nodes in the onset of a  global cascade by conducting counterfactual simulations in which false-active nodes are forcibly eliminated.
We will then compare the simulated results with and without false-active nodes.
}
The procedure of simulations \emph{without} false-active nodes is as follows:
\begin{enumerate}
     \item For given $\langle k\rangle$ and $N$, generate an \ER network with a common connecting probability $\langle k\rangle/(N-1)$. 
     Select seed nodes at random so that there are $\lfloor \rho_0 N\rfloor$ active nodes. The other nodes are inactive at $t=0$.
     \item Let ${\bm{v}}(t+\Delta t)$ denote the set of a fraction $\Delta t\in (0,1)$ of nodes selected uniformly at random at time $t+\Delta t$.
     Update the states of nodes in ${\bm{v}}(t+\Delta t)$ according to the activation rule \eqref{eq:threshold_F}. We set $\Delta t=0.01$.  
     
     \item Loop over all active nodes in ${\bm{v}}(t+\Delta t)$ and examine whether the deterministic threshold condition $\qtrue>\theta$ is satisfied or not. If there are false-active nodes, the nodes are deactivated.
     \item Reset $t$ as $t+\Delta t$.
     \item Repeat steps 2--4 until convergence, where no nodes change their states.  
     \item Repeat steps 1--5 and take the ensemble average.
 \end{enumerate}
 
 To quantify the presence of false-active nodes in the dynamics of diffusion, we define the cumulative fraction of false-active nodes as 
\add{
\begin{align}
    \fa &=  \int_{0}^T dt\sum_k p_k \sum_{m,\,\bar{q}\leq \theta}\rho_{k,m}^1(t) \notag \\
    &\approx \sum_{\tau=0}^{\lfloor T/\Delta t\rfloor}\hat{\rho}^{1,{\textrm{false}}}(\tau \Delta t)\Delta t,
 \label{eq:cumulative_false_active}
\end{align}
where the first line is based on the AME solution, in which the sum of $\rho_{k,m}^1$ is taken over all $m$ such that $\bar{q}\leq\theta$} (i.e., the true threshold condition is not satisfied). 
The integral is approximated by the sum in the second line where $\hat{\rho}^{1,{\rm{false}}}(\tau\Delta t)$ denotes the simulated fraction of false active nodes at $t=\tau\Delta t$ for $\tau=0,1,\ldots,\lfloor T/\Delta t\rfloor$.
$T$ is the time to convergence.
 Note that since seed nodes are initially active while their neighbors are all inactive, the fraction of false-active nodes at $t=0$ is $\rho_0$. 
 We assume that $\rho_0$ is small enough.
 For a sufficiently large $T$, the value of $\Theta$ that would be attained when there is no contagion, denoted by $\Theta_0$, is given by
 \begin{align}
    \Theta_0 &\approx \rho_0\Delta t + (1-\Delta t)\rho_0\Delta t + (1-\Delta t)^2\rho_0\Delta t+\cdots \notag \\
    & = \rho_0.
    \label{eq:theta_0}
 \end{align}
The first line reflects the updating process in which a fraction $\Delta t$ of all nodes change their states. 
In this hypothetical case where there is no contagious effect, all updating seed nodes will surely change the state from false active to true inactive.  
The fraction of false-active nodes therefore decays at a rate of $1-\Delta t$.
Eq.~\eqref{eq:theta_0} indicates that the cumulative fraction of false-active nodes in the absence of contagion is given by $\rho_0$.
Therefore, we normalize $\Theta$ by subtracting $\Theta_0\approx\rho_0$ to eliminate the direct influence of seed nodes.
The normalized $\Theta$, denoted by $\bar{\Theta}$, is then given by
\begin{align}
    \bar\Theta \coloneqq  \Theta - \rho_0.
\end{align}

Recall that a rise in $\lambda$ generally expands the cascade region (Figs.~\ref{fig:phase_diagram}b and \ref{fig:misperception_heatmap}). 
In fact, there is a certain value of $\lambda$ at which both $\rho^1$ and $\bar\Theta$ suddenly increase (Fig.~\ref{fig:cumulative_false_active}b), indicating that a small change in individuals' \namelambda   may cause a sudden shift to the opposite equilibrium.
Although such an increase in $\bar\Theta$ could be observed if there is a rise in $\rho^1$, our counterfactual simulation shows that a global cascade would not occur even above the critical value of $\lambda$ if false-active nodes were absent (blue triangles in Fig.~\ref{fig:cumulative_false_active}b).
This observation leads us to conclude that misperception can be a \emph{cause} of global cascades rather than an outcome.
In other words, social contagion may arise in a \emph{self-fulfilling} manner: a group of nodes that are activated due to misperception may initiate a global cascade, suggesting that the initially ``wrong'' decision to be active could turn out to be ``correct'' ex post.

Fig.~\ref{fig:cumulative_false_active}b also indicates that there is a certain degree of accommodation, denoted by $\widetilde{\lambda}$, at which the cumulative share of false active nodes is maximized.
When $\lambda$ is small, active nodes are scarce, and accordingly there are few false-active nodes.
When $\lambda$ is close to one, on the other hand, a global cascade surely occurs, and most nodes become true-active while there are only a few false-active ones.
This implies that the fraction of false-active nodes will be maximized between the two extremes.   


\subsection{State reversals}\label{sec:state_reversal}

 In the threshold model without uncertainty, it is known that nodes monotonically change their state from inactive to active, but not vice versa~\cite{Gleeson2007,kobayashi2022EconTheo}. 
 However, in the presence of uncertainty, nodes may revert their states from active to inactive (i.e., \emph{deactivation}), depending on the size of the signal they observe.

 To quantify the frequency of state reversals, we define $\mathcal{R}$ as the cumulative fraction of nodes that have been deactivated:
\begin{align}
    \mathcal{R} &=  \sum_{\tau=0}^{\lfloor T/\Delta t\rfloor-1}\frac{\big|\{i:\, s_i((\tau+1)\Delta t)- s_i(\tau\Delta t)<0\}\big|}{N}, \label{eq:reverse_fraction}
\end{align}
where $s_i(t)\in\{0,1\}$ denotes the state of node~$i$ at time $t$.
$|\{i:\, s_i((\tau+1)\Delta t)- s_i(\tau\Delta t)<0\}|/N$ denotes the fraction of nodes that reverted their state from active to inactive at time $(\tau+1)\Delta t$.

\begin{figure}[tb]
    \centering
    \includegraphics[width=7.5cm]{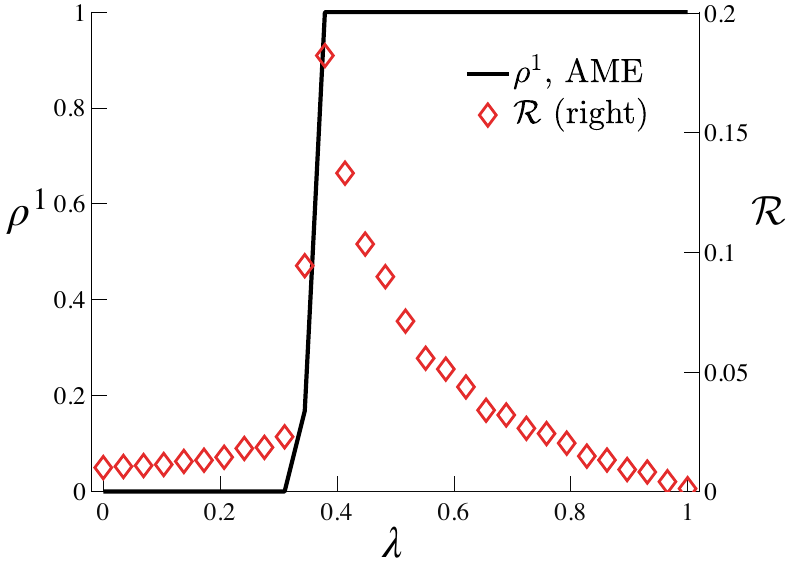}
    \caption{Steady-state fraction of active nodes and the cumulative fraction of nodes that reverted the states, $\mathcal{R}$.
    We set $N=10^4$, $\langle k\rangle =8$, $\theta=0.18$, $\beta=5$, $\rho_0=0.01$. Average is taken over $10,000$ runs.
    }
    \label{fig:reversal_lambda}
\end{figure}

We find that the reversal rate $\mathcal{R}$ drastically rises near the critical point above which a global cascade can occur (Fig.~\ref{fig:reversal_lambda}) and converges to $0$ as $\lambda\to 1$.
When $\lambda$ is less than the critical value, $\mathcal{R}$ takes values close to zero because the majority of nodes rarely become active.
When $\lambda$ is larger than the critical value, on the other hand, the majority of nodes become active sooner or later. 
The reversal rate decreases as individuals become more accommodative, because an increase in $\lambda$ makes it more likely that the fraction of active neighbors will increase monotonically during the contagion process.
This suggests that state reversals are most frequent near the critical point where there is a strong tension between activation and deactivation.

\add{
\section{Sequential Bayesian update}\label{sec:sequential_update}

So far, we have assumed that the prior  $f(q)$ is always unbiased (i.e., $\mathbb{E}(q)=\qtrue$), and the posterior $f(q|\mo)$ is also unbiased for a sufficiently large $\beta$ (i.e., $\lim_{\beta\to\infty}\mathbb{E}(q|\mo)=\qtrue$).
While the assumption of unbiasedness allows us to quantify the pure effect of uncertainty, the cost is that the prior parameter $\alpha$ is contingent on the true value $\qtrue$ (Sec.~\ref{sec:degree_uncertainty}).  
In this section, we instead consider a sequential Bayesian updating scheme in which the posterior obtained in the $(t\!-\!1)$-th update is used as the prior in the $t$-th update.
The updating rule is given by
\begin{align}
    f_t(q|\{\mo_t\}) =\frac{f(\mo_t|q)f_{t-1}(q|\{\mo_{t-1}\})}{\int_{0}^1f(\mo_t|q)f_{t-1}(q|\{\mo_{t-1}\})dq},
    \label{eq:sequential_update}
\end{align}
where $\{\mo_t\}\coloneqq (\mo_1,\ldots,\mo_t)^\top$ denotes the history of observed signals up to the $t$-th update. 
As before, we specify $f(\mo|q) = \binom{k}{\mo}q^\mo (1-q)^{k-\mo}$, and the initial prior is given by $f_{0}(q|\mo_{0})=f(q)$.

We assume that the prior in the initial update, $f(q)$, is specified as ${\rm Beta}(a,b)$ for $a,b>0$.
The posterior in the $t$-th update is then given by ${\rm Beta}(a+\sum_{\tau=1}^t\mo_\tau,b+tk-\sum_{\tau=1}^t\mo_\tau)$.
Here, we determine the parameter $a$ so that the prior mean in the initial update (i.e., $a/(a+b)$) is equal to the seed fraction $\rho_0(>0)$ for a given $b$, which gives $a = b\rho_0/(1-\rho_0)$.
The initial prior mean is thus given by
\begin{align}
    \mathbb{E}(q) &= \rho_0, 
\end{align}
Note that while this updating scheme appears natural in that the current prior is based on the information available at the time of updating, the prior mean is generally biased (i.e., $\rho_0\ne \qtrue$).  
This suggests that a cascade may occur due to the biased prior belief rather than the uncertainty effect.
The posterior mean leads to
\begin{align}
    & \mathbb{E}(q|\{\mo_t\}) = \frac{\rho_0 b +(1-\rho_0)\sum_{\tau=1}^t\mo_\tau}{b+(1-\rho_0)tk}. \label{eq:ave_q_sequential}
\end{align}
We have
$\lim_{t\to\infty}\mathbb{E}(q|\{\mo_t\})=\lim_{t\to\infty}\sum_{\tau=1}^t\mo_\tau/(tk)$, meaning that in the limit of large $t$, the posterior mean will be represented by the asymptotic average of realized signals.
Note that for any $t>0$, the average signal term  $\sum_{\tau=1}^t\mo_\tau/(tk)$ in Eq.~\eqref{eq:ave_q_sequential} also becomes dominant as $b\to 0$.
Therefore, the lower the parameter $b$, the faster the posterior mean will approach its asymptotic average.
In contrast, we have $\lim_{b\to\infty}\mathbb{E}(q|\{\mo_t\})=\rho_0$, meaning that
individuals will become less responsive to the observed signals as $b$ increases.

\begin{figure}[tbh]
    \centering
    \includegraphics[width=8.6cm]{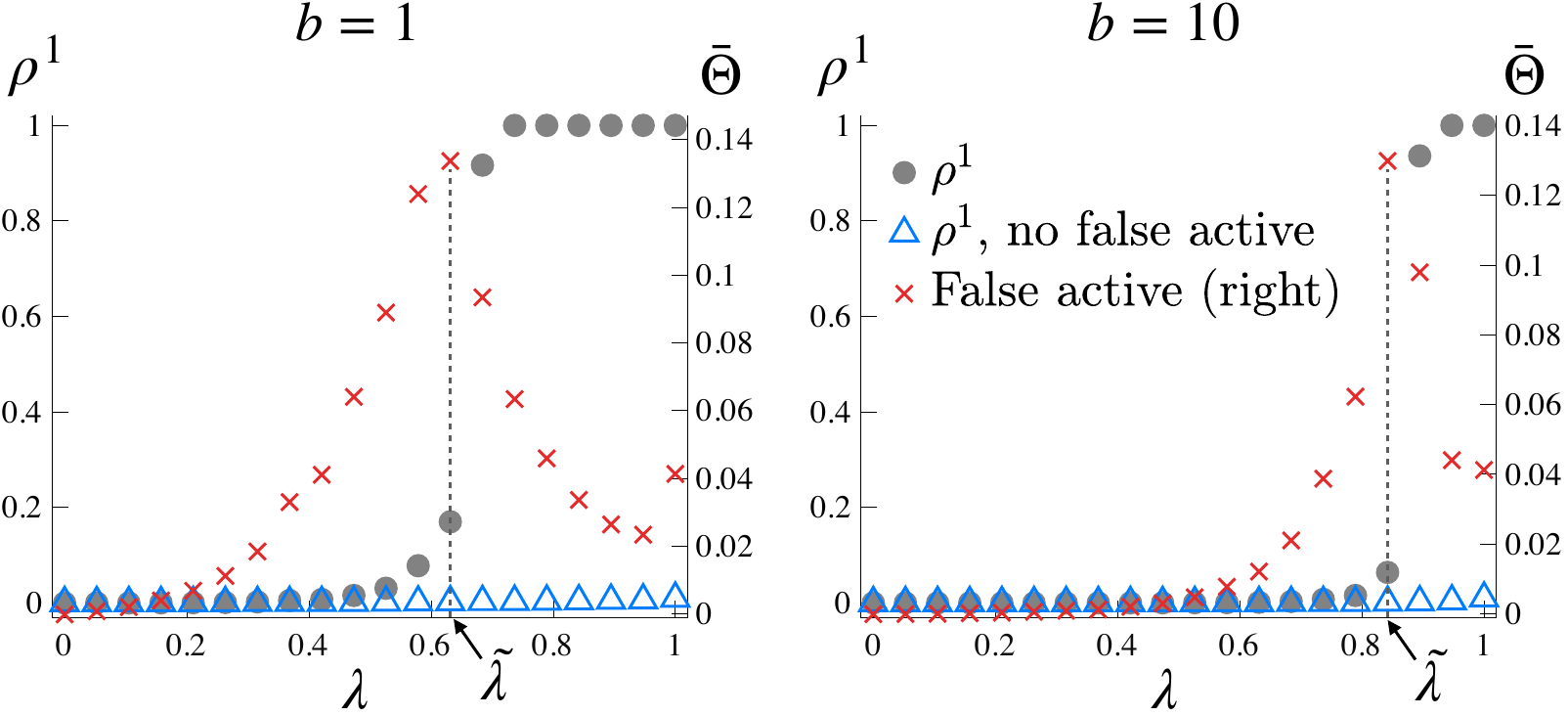}
    \add{
    \caption{
    Simulated cascade size under sequential Bayesian updating.
    The initial prior $f(q)$ is given by ${\rm Beta}(a,b)$, where $a=b\rho_0/(1-\rho_0)$.
    See the caption of Fig.~\ref{fig:cumulative_false_active} for the parameter settings.}
    }
    \label{fig:sequential_rho1}
\end{figure}

Fig.~7 shows simulated cascade sizes against $\lambda$.
As we saw in Fig.~\ref{fig:cumulative_false_active}b, we observe a sudden transition to a cascade region at a certain accommodation level (black circle).
The increase in $\bar{\Theta}$ (red cross) also indicates that a global cascade can occur only if there are a sufficient number of false active nodes.
In fact, the value of $\lambda$ at which the fraction of false active nodes reaches its maximum, denoted by $\widetilde{\lambda}$, coincides with  the critical point at which the fraction of active nodes increases drastically.
The critical point and $\tilde\lambda$ shift to the right as the prior parameter $b$ increases (Fig.~7, \emph{right}).

}

\section{Discussion}

In this paper, we developed a threshold model of cascades in the presence of noisy information. 
Each individual infers the distribution of the fraction of active neighbors based on noisy signals and uses the inferred distribution to decide whether to be active or not.  
The analysis showed that misperception due to observation uncertainty could promote or inhibit the spread of social contagion.

There are several issues left for future research.
First, while we considered observational uncertainty about the state of neighbors, there may be other types of uncertainty, such as the quality/reliability of the information.
In recent years, the spread of misinformation has been one of the critical social problems~\cite{ferrara2020misinformation,pennycook2022accuracy}; therefore, it is worthwhile to investigate how the quality of information might affect the spreading dynamics.
Second, we considered static networks and the cascade dynamics on them.
The model could be extended to study cascades in temporal networks, where the role of uncertainty is intrinsically important since the contact history of neighbors is generally stochastic.
\add{
Third, while we analyzed \ER random networks, more realistic network models having a scale-free degree distribution and/or a community structure need to be examined to understand the actual dynamics of complex contagion. 
}
In these respects, the current model should be considered as a first step toward a better understanding of cascades under uncertainty.

\section*{Acknowledgments}
Kobayashi acknowledges financial support from JSPS KAKENHI\ 20H05633 and 22H00827.

%


\end{document}